\documentclass[final,5p,times,twocolumn]{elsarticle}

\usepackage{graphics}
\usepackage{multirow}
\usepackage[usenames]{color}

\usepackage{amssymb}
\usepackage{amsmath}

\journal{International Journal of Engineering Science} 

\begin{document}

\begin{frontmatter}

\title{Dynamic bridging of proppant particles in a hydraulic fracture}


\author[SKT,IFZ]{I.A.~Garagash}
\author[SKT]{A.A.~Osiptsov}
\author[SKT]{S.A.~Boronin}
\ead{a.osiptsov@skoltech.ru}

\address[SKT]{Skolkovo Institute of Science and Technology (Skoltech), 3 Nobel Street, 143026, Moscow, Russian Federation}
\address[IFZ]{Institute of Physics of the Earth, Russian Academy of Science, Bolshaya
Gruzinskaya str., 10-1, 123242, Moscow, Russian Federation}

\begin{abstract}
This work is focused on the development of a dynamic criterion for the arching and bridging of spherical particles in a 3D suspension flow through a channel with plane walls. Elasticity of the particles and the channel walls are taken into account. The carrier fluid is viscous and incompressible. Bridging occurs under the balance of the hydrodynamic force exerted from the fluid on the particles and the friction force exerted from the walls on the particles. The 3D motion of particles in fluid is analyzed by means of direct numerical simulation. The bridging criterion is formulated as a domain on the plane in terms of the two nondimensional parameters: the particle size to channel width ratio and the flow velocity. For each scaled particle diameter there is a range of critical velocities, in which bridging occurs. Various geometrical configurations are considered: three and four particles across the slot. Stability of the bridge is studied. The dynamic bridging criterion is different from the earlier purely kinematic criteria, which were formulated in terms of the particle-to-channel width ratio only. The bridging criterion is implemented into the 2D width-averaged lubrication model of suspension flow through a plane channel, and illustrative simulations are conducted. Application is for proppant transport in hydraulic fractures.
\end{abstract}

\end{frontmatter}

\section{Introduction}
\label{Intro}

This paper presents the results of a research into the dynamic bridging criterion for spherical particles transported in a viscous flow through a plane channel. One of the key applications is bridging of proppant particles during suspension transport in a hydraulic fracture \cite{Osiptsov2017review}. Bridging is dangerous for hydraulic fracturing operations, when it occurs in the near-wellbore area, as it practically stops the injection of suspension and leads to screen-out. Hydraulic fracturing operation is typically designed with the help of simulators based on coupled mathematical models of the process of fracture growth and suspension placement in the fracture~\cite{Desroches,Barbati,Osiptsov2017review}. In a complex formulation with account for variety of relevant phenomena, the model of proppant transport in a fracture is based on 2D width-averaged lubrication approach within the two-continua two-speed formulation for suspension flow~\cite{BoroninOsiptsov2010} with closure relations for particle settling velocity, suspension rheology, sub-models for tip screenout \cite{Chekhonin, Dontsov2015}, effects of shear stress on frac tip growth~\cite{Mishuris2017} and effects of cohesion on fracture tip propagation~\cite{Garagash2018}, to name a few. Global geomechanics framework for fracture growth propagation is now ranging from enhanced P3D~(e.g., \cite{DontsovEP3D}) to Planar3D (also including the poroelastic effects, e.g.~\cite{Golovin}) and full 3D for the main hydraulically induced fracture, with a number of emerging approaches to modelling the fracture network in shales. More detailed description can be found in recent reviews~\cite{Detournay, Osiptsov2017review}.

In this paper, we focus on a particular effect, namely, bridging or arching, which is caused by geometric constraints. In a pressure driven flow of suspension through a fracture (approximated by a narrow channel with rough walls), interactions between particles and with the walls result in particles lagging behind the flow and slowing down, which will eventually result in the increase of the mean particle volume fraction. When the fracture width gets small enough for the bridging criteria to be met, this process eventually leads to the concentration reaching maximum packing limit. At some stage, however, it is also possible
that even before reaching $C_{max}$, the particles may literally become lodged between the slot faces. Because the particles do not re-arrange in the form of a regular train along the fracture, they get locked across the slot, forming a "bridge" or an "arch" between the faces~\cite{Osiptsov2017review}. This bridge, once formed, is usually kept in place by the balance between the hydrodynamic force exerted from the flowing fluid and the contact stresses from face to face through the particles. Also important are the roughness of the particles and fracture faces. Once a bridge is formed, it usually results in the arrest of the suspension transport upstream this point, which may result in a screenout and a complete stop of the pumping process during hydraulic fracturing. 

The most commonly used model for bridging consists of defining a certain threshold width, at which the particles form a bridge or an arch. If in any part of the fracture the width is equal or less than this critical width $w^*$, the proppant should not be allowed to flow through it. The work \cite{Van-der-Vlis} determined experimentally the bridging factor $b=w^*/d=2.6$, which
is the criterion currently used by some commercial simulators by default. For low proppant loading, the bridging factor was found to be 1.8. In the industry, it has become customary to use this simplified criterion of bridging in terms of the bridging factor taken as a constant from the interval $w^*/d\in [2.5-3]$ \cite{GuDesroches, Dontsov2014a}.

More sophisticated criteria take into account experimental observations \cite{GruesbeckCollins}, which demonstrate that the critical width is also dependent upon the concentration of proppant upstream of the bridging point. Although the original criterion of \cite{GruesbeckCollins} was developed for modeling bridging at perforations, there was a modified version of bridging criterion in a fracture, which demonstrates \cite{MackWarpinski, Osiptsov2017review} 
$$
w^* = \mathrm{min}\left[b,1 +\frac{C_p}{0.17}(b-1)\right]d
$$
%
%
where the bridging factor $b=2.5$ by default, $d$ is the particle diameter, and $C_p$ is the particle volume fraction in flowing suspension. 

The effects of fibers on bridging have been studied in \cite{Vasudevan, Vasudevan1} based on the partitioning of energy approach following Bagnold's theory. Flow energy imparted to the moving slurry is partitioned into useful shear work, kinetic energy of particles, and a portion lost in dissipative solid-body interactions. A modified bridging criterion was proposed in the form ${\rm B}_{cr}\sim T_{EP}T_{B}T_{FC}$, where $T_i$ are energy spent for energy partitioning, blockage, and particle-fibers ineractions, respectively. Closures for $T_i$ were expressed via volume fraction of fibers $C_f$ based on phenomenological considerations, though we think fibers crowding parameter $N=(2/3)C_f (l_f / d_f)^2$ \cite{Paper} is a better measure of fibers impact than pure fibers volume concentration. 
A recent experimental study of bridging in slots with tapered walls can be found in \cite{Ray}, where the key finding was that for slots with smooth walls $b=w^*/d\approx 1$.

As we noted above, on the fracture scale, bridging and packing in application to tip screenout (TSO) was studied in \cite{Chekhonin, Dontsov2014, Dontsov2015}. The work \cite{Chekhonin} split the entire fracture into the flowing suspension region $C_p~<~C_{max}$ and the porous medium of packed proppant $C_p=C_{max}$, where the Darcy law applies. KGD model was used for fracture propagation, and a more complex than Carter's relation was used for leak-off. In turn, the works of \cite{Dontsov2014, Dontsov2015} utilized frictional rheology of \cite{Boyer} to study 3D slurry flow in a fracture and transition to packing, also including transition to Darcy's flow in packed region. Detailed comparison with concurrent works of \cite{Lecampion2014} are given.

This paper is organized as follows. Section 2 presents a detailed description of the model for dynamic bridging for loose packing (Sec. 2.2) and close packing (Sec. 2.3) configurations. The implementation of the resulting dynamic bridging criterion into the 2D proppant transport model in a research simulator is given in Sec. 3. Discussion (Sec. 4) and concluding remarks (Sec. 5) are provided at the end.

\section{Dynamic bridging criterion from 3D direct numerical simulation}

As discussed above the motion of particles in the flowing suspension in a hydraulic fracture may be retarded and arrested due to inter-particle interaction and friction with the walls thus resulting in the formation of a bridge or an arch across the fracture. The existing models of bridging in a fracture trace back to the original works on bridging in perforation tunnels~\cite{GruesbeckCollins}, which are circular tubes rather than place channels. Theory of particle bridging in plane channels needs further development, though experimental work is progressing and there is a number of recent studies in the lab (e.g.,~\cite{Ray}). Thus, in order to investigate the phenomenon of bridging one needs to analyze the conditions of particles interaction with the fracture walls and between each other to propose adequate models for phenomenological description of particle motion and bridging formation.

\subsection{Numerical modeling of interaction of particles with fracture walls}

\begin{figure}[h]
\centering
\includegraphics[width=6cm]{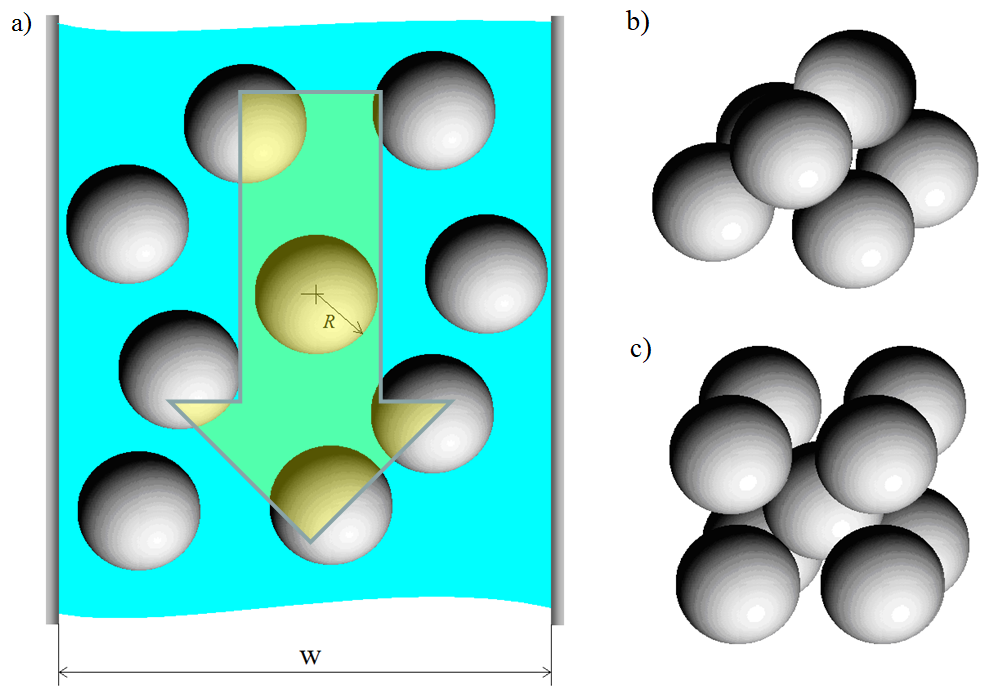}
\caption{Possible configurations of particles across the fracture}
\label{Fig1}
\end{figure}

We consider the process of bridging formation in a fracture. When the grain radius is significantly smaller than the fracture width Fig.~\ref{Fig1}, {\it a}, one may expect the motion of particles in fluid through the fracture to be unconstrained. As the fracture width decreases, from 4 to 3 grains may be placed over the fracture width, so the following condition is satisfied $w<4R$. In this case, the particle packing may take different forms and configurations. Two possible configurations are shown in Fig.~\ref{Fig1}, {\it b} and {\it c}. 

\begin{figure}[!htb]
\centering
\includegraphics[width=6cm]{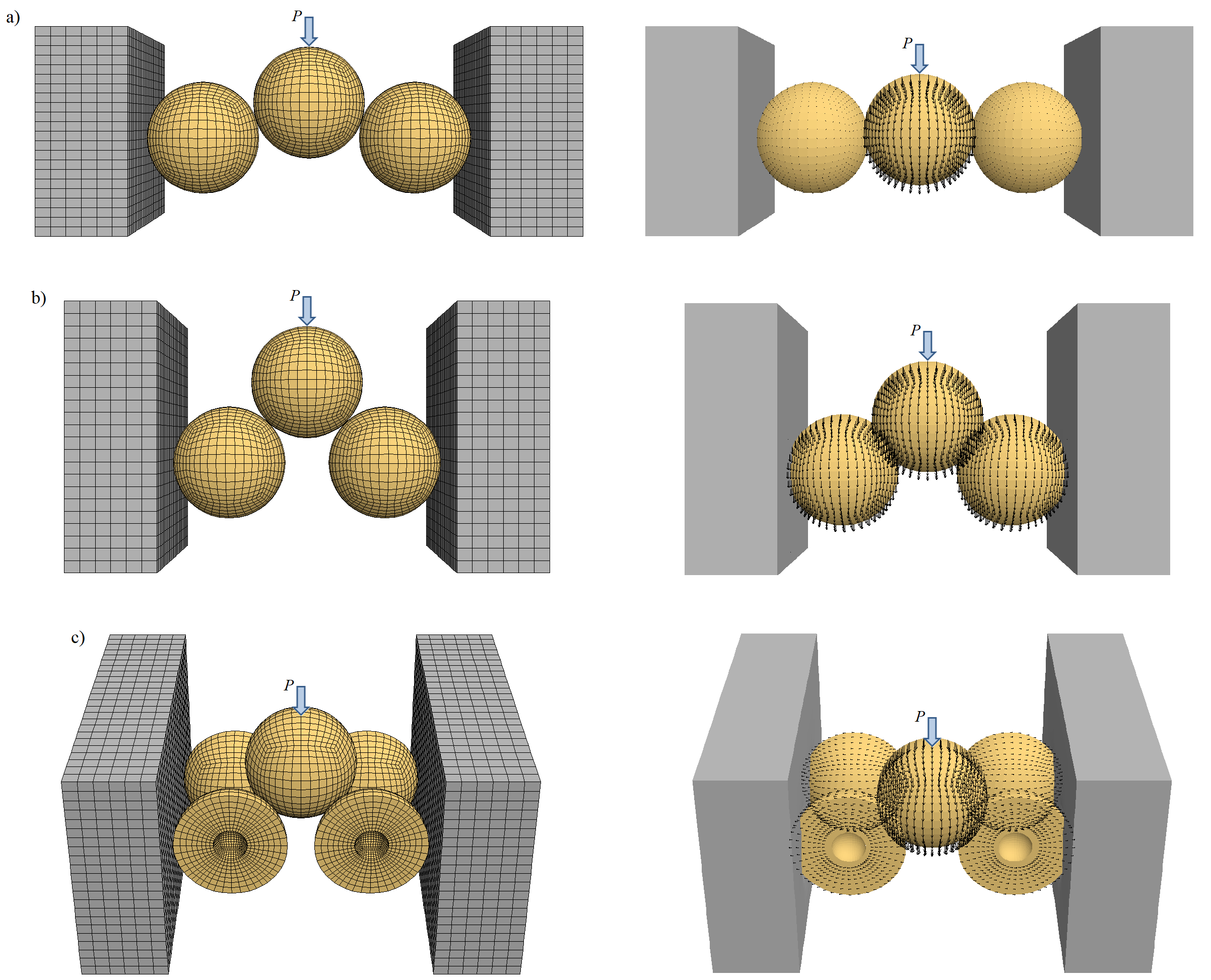}
\caption{Velocity distributions in elastic system "proppant-fracture": unconsolidated packing of grains $W/d=2.932$ ({\it a}), unconsolidated packing of grains $W/d=2.414$ ({\it b}), and close packing of grains $W/d=2.113$ ({\it c}).}
\label{Fig2}
\end{figure}

According to Fig. ~\ref{Fig1}, {\it b}, we will study the interaction of three elastic spherical proppant grains in the fracture (Fig. ~\ref{Fig2}, {\it a}).

\subsection{Conditions for particle bridging in a loose packing}
\begin{figure}[!htb]
\centering
\includegraphics[width=6cm]{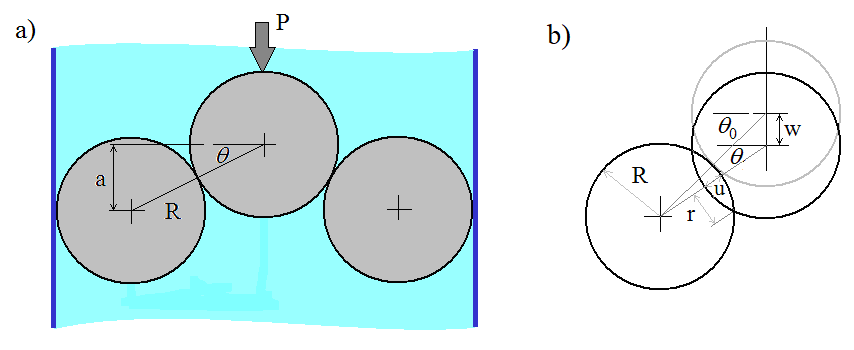}
\caption{Proppant grains ({\it a}) and the geometry of contact ({\it b}).}
\label{Fig3}
\end{figure}

 We consider a loose (unconsolidated) packing of particles. According to Fig. \ref{Fig1}, {\it b}, we study the interaction of three elastic spherical granules of proppant in a fracture (Fig. \ref{Fig3}, {\it a}). The geometry of the packing is characterized by the following relation:
\begin{equation} 
W/d=1+2\cos\theta^{0}
\label{Eq1}
\end{equation}

Based on consideration of the geometry of contact, we obtain the following relations which relate the vertical displacement of the central sphere $w$ with the variation of the angle $\theta$ and the displacement of spheres $u$,
\begin{equation}
\theta = \arctan\left(\frac{2R\sin\theta_0-w}{2R\cos\theta_0} \right), \quad u=\frac{2R(\cos\theta-\cos\theta_0)}{\cos\theta} 
\label{Eq2}
\end{equation}

In what follows we consider the equilibrium of spheres (Fig. \ref{Fig2}). 
\begin{figure*}[h]
\centering
\includegraphics[width=15cm]{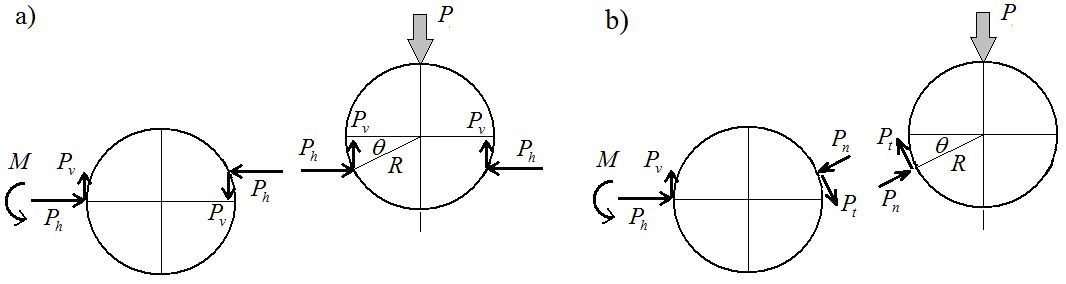}
\caption{Forces acting at the contacts between grains.}
\label{Fig4}
\end{figure*}

According to Fig. \ref{Fig3}, {\it a}, vertical force $P_{\nu}$, horizontal force $P_h$, and the moment $M$ at the fracture wall are related by the formulas:
\begin{equation}
P_v=\frac 12P,\quad P_vR[1+\left(1-\frac{u}{2R}\right)\cos\theta]-P_hR\left(1-\frac{u}{2R}\right)\sin\theta-M=0
\label{Eq3}
\end{equation}
The rolling friction related to the moment $M$ is not considered in what follows. 
\begin{figure*}[h]
\centering
\includegraphics[width=18cm]{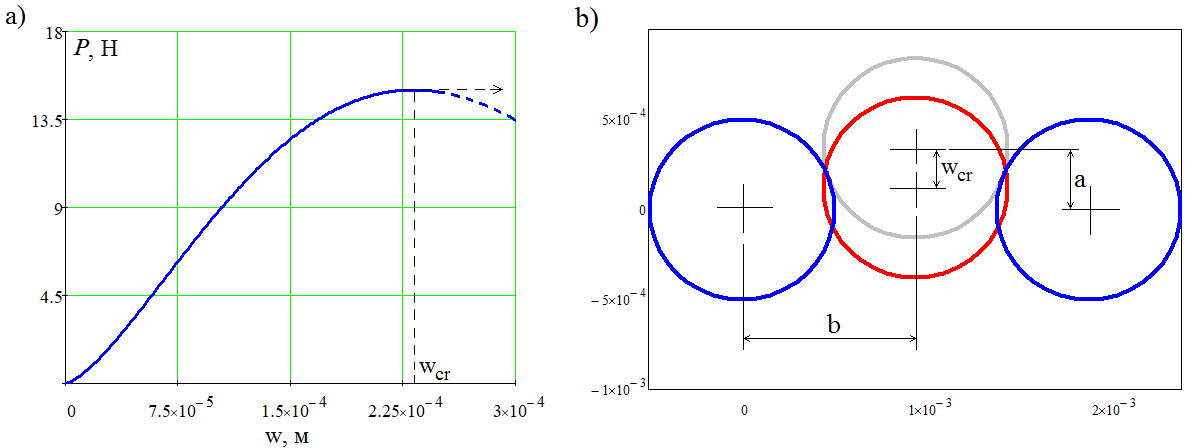}
\caption{Dependence of the force $P$ on the displacement ({\it a}); The position of the central particle at the initial time instant (grey contour) and in the moment of breakthrough (red contour) ({\it b}).}
\label{Fig5}
\end{figure*}

    Forces $P_{\nu}$ and $P_h$ can be expressed via normal and tangential forces at the contact surface (Fig. \ref{Fig4}, {\it b}) as follows:  
\begin{equation}
P_v=P_n\sin\theta + P_t\cos\theta, \quad P_h = P_n \cos\theta - P_t \sin\theta  
\label{Eq4}
\end{equation}

According to the Hertz solution \cite{Bazant}, the displacement $u$ and the normal force $P_n$ at the contact surface are related as:
\begin{equation}
P_n=\frac{2G}{3(1-\nu)}
\label{Eq5}
\end{equation}
where $G$ -- Young's modules, and $\nu$ -- Poisson's ratio.

Since it is assumed that the central sphere is moving slowly, overcoming the friction forces under the action of the increasing force $P$, the tangential force $P_t$ can be written in the form:
\begin{equation}
P_t=\alpha P_n
\label{Eq6}
\end{equation}

where $\alpha$ is the dry friction coefficient. 

Using relations (\ref{Eq3})--(\ref{Eq6}), we obtain the expression for the force $P$, which acts to embed the sphere into the spacing between the other two grains, in the form:
\begin{equation}
P=\frac{4G}{3(1-\nu)}\sqrt{2Ru^3}(\sin\theta + \alpha\cos\theta)
\label{Eq7}
\end{equation}

	We consider the grains with the Young's modulus $G=2\cdot 10^9$ MPa, Poisson's ratio $\nu=0.3$ and the friction factor $\alpha=0.2$. The grains have the radius $R=5\cdot 10^{-5}$ m and are initially located under $P=0$ at the angle $\theta_0$ with respect to each other. 
    
    The simulation shows that, as the central particle is penetrating between the two granules, the load $P$ increases, and then, when the displacement reaches a critical threshold $w=w_{cr}$, the load decreases and the central particle moves through the constriction between the other two particles (Fig. \ref{Fig5}, {\it a}).
   
\begin{figure}[!htb]
\centering
\includegraphics[width=8cm]{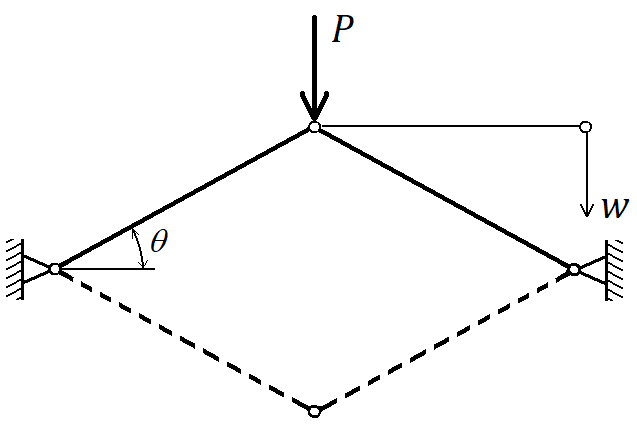}
\caption{The von Mizes truss.}
\label{Fig6}
\end{figure}

In this case the angle reaches the critical value of $\theta_{cr}=6.8^0$. The position of the central particle is shown by the red contour in Fig. \ref{Fig5}, {\it b}. One may derive the conjecture that the particles behave as a system with jumps (or a system with non-adjacent equilibrium positions), where the transition to the new equilibrium position occurs due to the accumulated elastic energy in the system. In this sense, the behavior of the grains is similar to the Mizes truss (Fig. \ref{Fig6}). 

The von Mises truss is an example of a globally unstable system~\cite{Mises, Bazant}. When the load reaches a certain maximum threshold value, the system "jumps" into a new equilibrium position shown in Fig.~\ref{Fig6} by the dashed line.

Note that the motion of the central sphere occurs while the side grains touching the walls are immovable, and the load $P_t$ at the points of contact of the central sphere exceeds the friction force. According to (\ref{Eq3})--(\ref{Eq4}), we introduce the following parameter $D$:
\begin{equation}
D=\left[ \frac{\alpha_w}{\cos\theta + 2\alpha_w\sin\theta}-(\alpha\cos\theta + \sin\theta\right]
\label{Eq8}
\end{equation}
 
 If the parameter $D>0$, then starting from the instant when the contact is formed at $\theta\ge\theta^0$, a bridge is formed across the fracture and the flow starts to decelerate. We will now find the interval $\theta_{max}\ge\theta\ge\theta_{min}$, where the bridging is formed. The threshold value of the angle, where deceleration begins, is $\theta_{min}=0$. According to (\ref{Eq8}), for deceleration to occur one need that the friction coefficient at the fracture wall is larger than the inter-particle friction coefficient, i.e. $\alpha_w>\alpha$ and $D=\alpha_w-\alpha$. The angle $\theta_{max}$ will be found from the condition $D=0$. Using expression (\ref{Eq8}), we find:
\begin{equation}
\theta_{max}=\arcsin\sqrt{-A-\sqrt{A^2-B}},
\label{Eq9}
\end{equation} 
where
$$
A=-\frac{1+2\alpha(\alpha-\alpha_w}{2(1+\alpha^2)(1+4\alpha_w^2)},\quad B=\frac{(\alpha_w-\alpha)^2}{(1+\alpha^2)(1+4\alpha_w^2)}
$$ 
 
 The existence of bridging in the identified range is bounded by instability. The angle $\theta_{cr}$, which marks the pull through of the central sphere in between the two side spheres, should satisfy the condition: 
\begin{equation}
\frac{\partial P}{\partial \theta}=0
\label{Eq10}
\end{equation} 
 
Substituting (\ref{Eq10}) into (\ref{Eq7}) and (\ref{Eq2}) after a number of transformations we obtain the expression for the critical angle:

\begin{equation}
\theta_{cr}\approx -\frac{2+\cos\theta_0}{6\cos\theta_0}\alpha +\sqrt{\left(\alpha\frac{2+\cos\theta_0}{6\cos\theta_0} \right)^2 + 2\frac{1-\cos\theta_0}{3\cos\theta_0}}
\label{Eq11}
\end{equation}

Thus relations (\ref{Eq9}) and (\ref{Eq11}) allow one to obtain the domain of existence for bridging depending on the friction coefficient $\alpha_w$ for a given value of the coefficient of friction between the grains $\alpha$. 

The bridge disintegration may also occur, if the mean normal stress in the contact zone
\begin{equation}
\sigma=\frac{P_n}{\pi r^2}=\frac{8G}{3\pi(1-\nu)}\sqrt{\frac{u}{2R}}\frac{1}{(2-u/2R)}
\label{Eq12}
\end{equation}
exceeds the limit of compression strength $\sigma_{fr}$.

From the condition $\sigma=\sigma_{fr}$ we find the expression for the threshold displacement:
\begin{equation}
u_{fr}=2R\left[\frac{M^2+4}{2}-\sqrt{\left(\frac{M^2+4}2\right)^2-4}\right]
\label{Eq13}
\end{equation}
where 
$$
M=\frac{8G}{3\pi(1-\nu)\sigma_{fr}}.
$$

The corresponding threshold force $P_{fr}$ has a form
\begin{equation}
P_{fr}=\frac{4G}{3(1-\nu)}\sqrt{2R u_{fr}^3}(\sin\theta_{fr}+\alpha\sin\theta_{fr})
\label{Eq14}
\end{equation}
where 
$$
\theta_{fr}=\arccos\left(\frac{\cos\theta}{1-u_{fr}/2R}\right).
$$

\begin{figure*}[!htb]
\centering
\includegraphics[width=15cm]{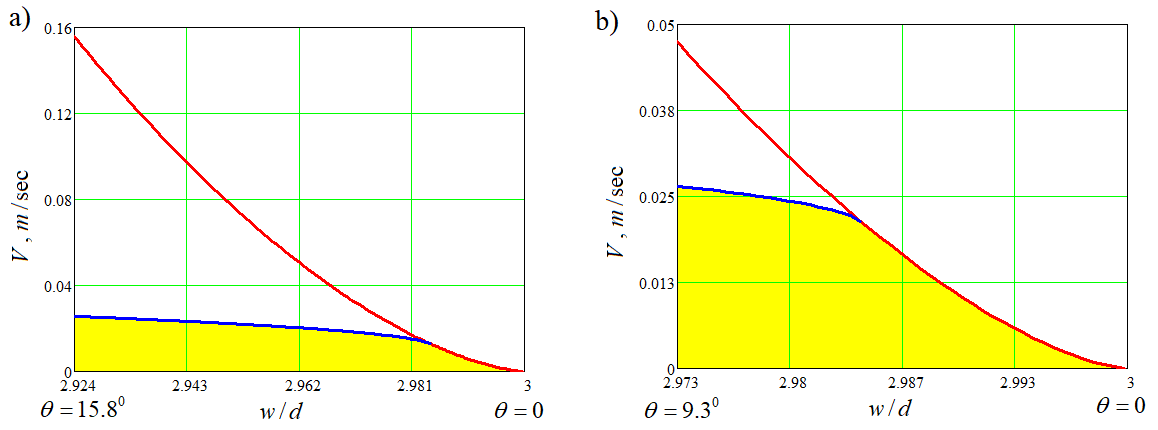}
\caption{Domains of existence for bridging (shown as yellow) for $\alpha_w=0.6$ in cases $\alpha=0.2$ ({\it a}) and $\alpha=0.35$ ({\it b}). On red curve, the central grain jumps through the side ones. On blue curve, the grains begin to crush.}
\label{Fig7}
\end{figure*}

The force $P$ depends on the velocity of fluid flow in the hydraulic fracture. Assume that the hydrodynamic force $P$ exerted from the fluid on the particle is governed by the Stokes expression $F_{St}$:
\begin{equation}
V=\frac{P}{6\pi \mu R}
\label{Eq12}
\end{equation}
where $\mu$ is the pure fluid viscosity.



The relations (\ref{Eq7}), (\ref{Eq9}), (\ref{Eq11}), and (\ref{Eq14}) together with the expression (\ref{Eq15}) allow us to obtain the domain of existence of bridging. Consider proppant grains of the radius $R=5\cdot 10^{-4}m$, Young's modulus $G=1\cdot 10^9$ Pa, Poisson's ratio $\nu=0.25$ and the compression strength $\sigma_{fr}=5\cdot 10^{7}$Pa. Fluid viscosity is $\mu=2.2\cdot 10^3$. Fig.~\ref{Fig7} shows the plots of velocity vs. fracture width to particle diameter ratio, where the domains of existence of bridging are shown. For $\alpha_w=0.6$, at $\alpha=0.2$ the domains of bridging are shown in Fig.~\ref{Fig7}, {\it a} (initial angles in the range $0\le \theta\le 15.8^{\circ}$), and at $\alpha=0.35$ -- in Fig.~\ref{Fig7}, {\it b} (initial angles in the range $0\le \theta\le 9.3^{\circ}$). For the domains of bridging existence thus identified, the bridge disintegration may occur both via the push (jump) of the central grain through the side ones (red curve in Fig.~\ref{Fig7}) and via the proppant grain crushing (blue curve in Fig.~\ref{Fig7}). These curves bound the domains of bridging in Fig.~\ref{Fig7}.

The analysis conducted so far allows one to derive a conclusion that there is a finite domain of bridge existence restricted in terms of both the maximum and minimum fluid velocity and in terms of the fracture-to-particle width ratio. In other words, in a gradually narrowing fracture there is the first threshold in terms of the fracture-to-particle width ratio $W/d$, when the bridge starts to form. With the further decrease in the fracture width, there is the second threshold in terms of the fracture-to-particle width ratio, where the bridge is no longer formed. This observation is confirmed by numerical experiments with loose proppant packings (Fig. 2, {\it a,b}). The same is true in terms of the velocity. With the increase in fluid velocity, there is the firs threshold below which the bridge does not form (the hydrodynamic force translating into wall friction is not sufficient to keep the bridge). Above this first threshold in terms of the velocity, the bridge starts to form. Then there is the second threshold in velocity, when the bridge disintegrates because the hydrodynamic force due to high fluid velocity is sufficient to push the central particle between the two side grains.
%
%
%
%
%
%
%
%
%
%
%
%

\subsection{Conditions of bridging in a close packing}

\begin{figure}[!htb]
\centering
\includegraphics[width=8cm]{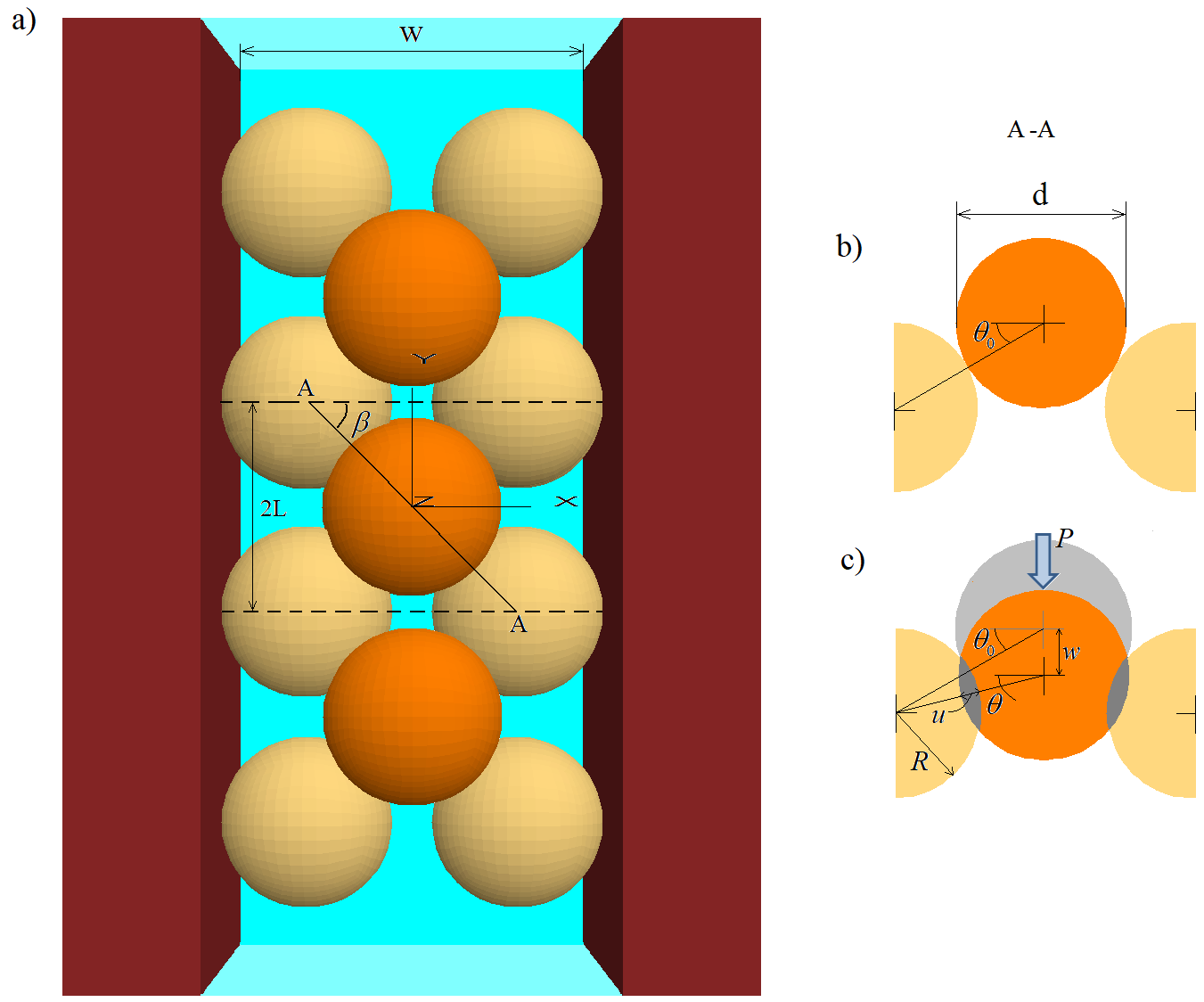}
\caption{Proppant grains ({\it a}) and the geometry of contact ({\it b}).}
\label{Fig10}
\end{figure}

We will now consider a close packing of proppant particles (Fig.~\ref{Fig10}). We assume that the suspension is dense, so the motion of particles in the cross-flow direction $y$ can be neglected and one can consider the subdomain shown in Fig.~\ref{Fig10} by the dashed line. 

The packing geometry satisfies the relations:
\begin{equation}
\frac{W}{d}=1+2\cos\theta^0\cos\beta,\quad \frac{L}{d}=\cos\theta^0\sin\beta
\label{Eq13}
\end{equation}

Considering the section A-A (Fig.~\ref{Fig10}, {\it a}) we will obtain the formulae, which relate the vertical displacement of the central sphere $w$ with the variation of the angle $\theta$ and the spacing between the spheres $u$:
\begin{equation}
\theta=\arctan\left(\frac{2R\sin\theta_0-w}{2R\cos\theta_0}\right),\quad u=\frac{2R(\cos\theta-\cos\theta_0}{\cos\theta}
\label{Eq14}
\end{equation}
where $\theta_0$ is the value of the angle at the present time instant. 

\begin{figure}[!htb]
\centering
\includegraphics[width=8cm]{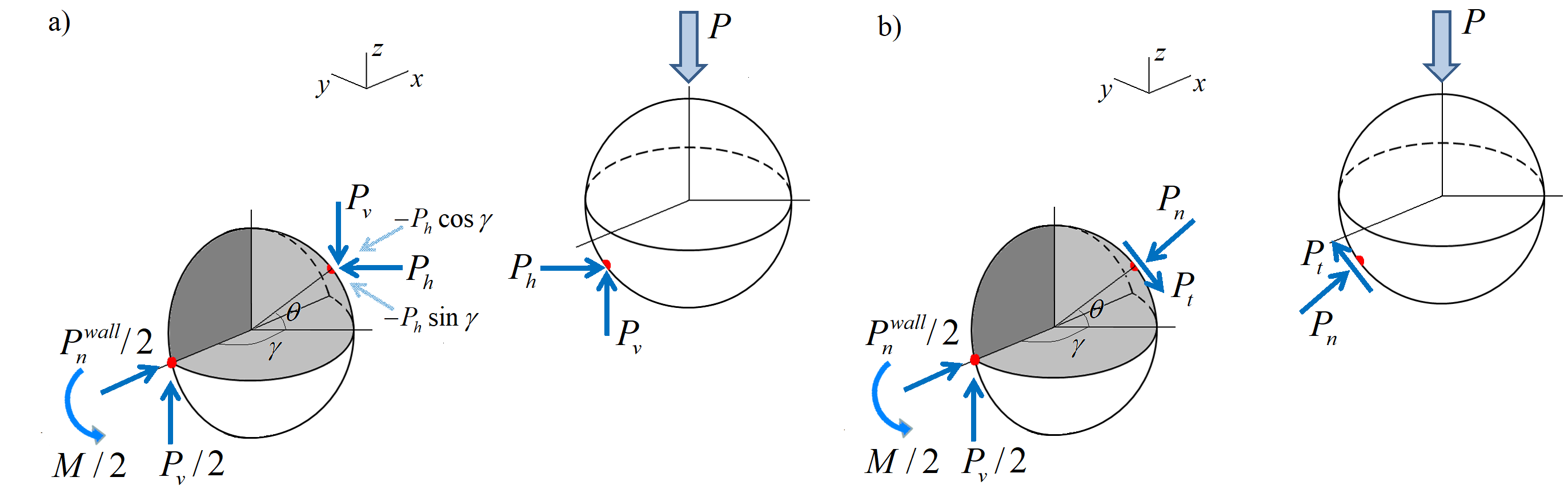}
\caption{Loads acting at the points of contact between the grains.}
\label{Fig11}
\end{figure}

In what follows we will consider equilibrium of the spheres. According to Fig.~\ref{Fig11}, {\it a}, the vertical $P_v$ and the horizontal $P_h$ loads and the moment $M$ are related by:
\begin{equation}
\begin{gathered}
P_v=\frac 1n P,\quad 2P_vR\left[ \left(1-\frac{u}{2R}\right)\cos\theta\cos\beta + 1\right]-{}\\
{}-2P_hR\left(1-\frac{u}{2R}\right)\cos\beta\sin\theta+M=0.
\label{Eq15}
\end{gathered}
\end{equation}

Forces $P_v$ and $P_h$ can be expressed via the normal $P_n$ and the tangential $P_t$ forces at the surface of contact (Fig.~\ref{Fig11}, {\it b}) by the following relations:
\begin{equation}
P_v=P_n\sin\theta+P_t\cos\theta,\quad P_h=P_n\cos\theta-P_t\sin\theta
\label{Eq16}
\end{equation}

As long as it is assumed that the central sphere is moving slowly overcoming the friction forces under the action of the increasing load $P$, the tangential load $P_t$ can be written as:
\begin{equation}
P_t=\alpha P_n,
\label{Eq17}
\end{equation}
where $\alpha$ -- the dry friction coefficient.

Using Eqs.~(\ref{Eq13}), (\ref{Eq13}), (\ref{Eq13}), the expression for the force $P$ which embeds the central sphere into the spacing between the grains can be represented in the following form:
\begin{equation}
P=\frac{8G}{3(1-\nu)}\sqrt{2Ru^3}(\sin\theta+\alpha\cos\theta)
\label{Eq18}
\end{equation}

\begin{figure}[!htb]
\centering
\includegraphics[width=8cm]{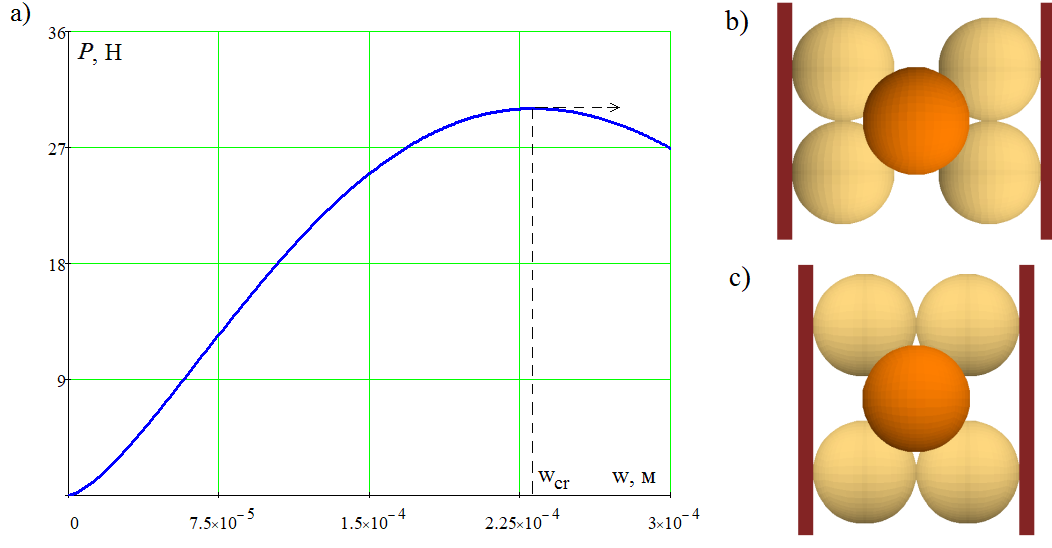}
\caption{Dependence of the load $P$ on the displacement $w$ ({\it a}); possible geometrical configurations of the central and side grains at the initial time instant ({\it b, c}).}
\label{Fig12}
\end{figure}

As in the previous case, we specify Young's modulus to be $G=2\cdot 10^9$ Pa, Poisson ratio $\nu=0.3$, and the friction coefficient $\alpha=0.2$. 
Let the particles having the radius $R=5\cdot 10^{-4}$ m be 
located at the initial distance at $P=0$ under the angle $\theta_0=20^0$ with respect to each other.
Then according to (\ref{Eq18}) we plot the increasing load $P$ (Fig.~\ref{Fig12}). Similar to the random loose packing the load is initially growing, 
and then after the displacement reaches the crtical threshold value $w=w_{cr}$ 
the particle abruptly moves through the spacing between the side grains (Fig.~\ref{Fig12}, {\it a}). Possible particle packings in this case are shown in Fig.~\ref{Fig12}, {\it b, c}.

The motion of the central sphere is realized while the side grains remain immovable. We note that it takes by a factor of two larger load to embed the central sphere between the side grains, as compared to the loose packing case (Fig.~\ref{Fig5}). It is possible under the condition that the vertical load $P_v$ is smaller than the friction at the wall $\alpha_wP_h$, i.e.:
\begin{equation}
P_v\le \alpha_wP_h\cos\beta,
\label{Eq19}
\end{equation}
where $\alpha_w$ -- the coefficient of friction at the fracture wall. 

Thus, when conditions (\ref{Eq17}) and (\ref{Eq19}) the particles decelerate and a bridge is formed. We introduce the parameter $D$ by the formula:
\begin{equation}
D=\left[\frac{\alpha_w\cos\beta}{\cos\theta+\alpha_w\sin\theta\cos\beta}-(\alpha\cos\theta+\sin\theta) \right] 
\label{Eq20}
\end{equation}

Then, the bridging is formed if $D>0$ starting from the instant of contact at $\theta\ge \theta^0$. The bridge forms if the ratio $w/d$ is in the interval between $(w/d)_{min}$ and $(w/d)_{max}$. The margin $(w/d)_{min}$ may be identified by 

\begin{equation}
\left(\frac{w}{d}\right)_{min}=1+2\cos\beta,
\label{Eq22}
\end{equation}
which corresponds to the condition $\theta_{min}=0$.

In order to find $(w/d)_{max}$, we will use the following expression:
\begin{equation}
\left(\frac{w}{d}\right)_{min}=1+2\cos\theta_{max}\cos\beta.
\label{Eq25}
\end{equation}

Using the condition $D=0$, we find:
\begin{equation}
\theta_{max}=\arcsin\sqrt{-A-\sqrt{A^2-B}}
\label{Eq26}
\end{equation}
where
$$
A=\frac{(\alpha_w\cos\beta-\alpha)(\alpha-\alpha_w\cos\beta)-0.5(1+\alpha\alpha_w\cos\beta)^2}{(\alpha-\alpha_w\cos\beta)^2+(1+\alpha\alpha_w\cos\beta)^2}
$$
$$
B=\frac{(\alpha_w\cos\beta-\alpha)^2}{(\alpha-\alpha_w\cos\beta)^2+(1+\alpha\alpha_w\cos\beta)^2}
$$

As in the case of loose packing, the existence of bridging in the interval identified is bounded by instability and possible crushing. The bridge, once formed, exists until the increasing fluid flow velocity breaks down the bridge by pushing the central particle in between the side grains. The critical angle $\theta_{cr}$ and the critical displacement of the grains $u_{fr}$ are determined by the relations (\ref{Eq11}) and (\ref{Eq11}), respectively.


The force which results in the central grain breakthrough:
\begin{equation}
P_{cr}=\frac{4G}{3(1-\nu)}\sqrt{2Ru^3}(\sin\theta_{cr}+\alpha\cos\theta_{cr})
\label{Eq25}
\end{equation}

is formed as a result of viscous flow around the grains at the velocity of:
\begin{equation}
V_{cr}=\frac{P_{cr}}{6\pi \mu R}
\label{Eq26}
\end{equation}

Here 
\begin{equation}
u_{cr}=\frac{2R(\cos\theta_{cr}-\cos\theta_0}{6\cos\theta_{cr}}
\end{equation}

The relation~(\ref{Eq25}) allows once to plot the curve of the critical flow velocity (when the bridge disintegrates) in the interval~(\ref{Eq22}).
 
For doing so we use the Stokes expression for the hydrodynamic force exerted on the grain:
\begin{equation}
V_{cr}=\frac{P_{cr}}{6\pi \mu R}
\label{Eq27}
\end{equation}

Bridging disintegration may also occur as a result of particle cracking in the contact zone. The normal stress at the contact zone 
\begin{equation}
\sigma_n=\frac{P_n}{\pi r^2}=\frac{4G\sqrt{u/2R}}{3\pi (1-\nu)(1-u/2R)}
\label{Eq28}
\end{equation}

should exceed the compression tensile strength $\sigma_{fr}$.
From the condition of $\sigma_{n}=\sigma_{fr}$ one may obtain the threshold value $u_{fr}$, and then the flow velocity $V_{fr}$ which breaks the grains. 

\begin{figure*}[!htb]
\centering
\includegraphics[width=15cm]{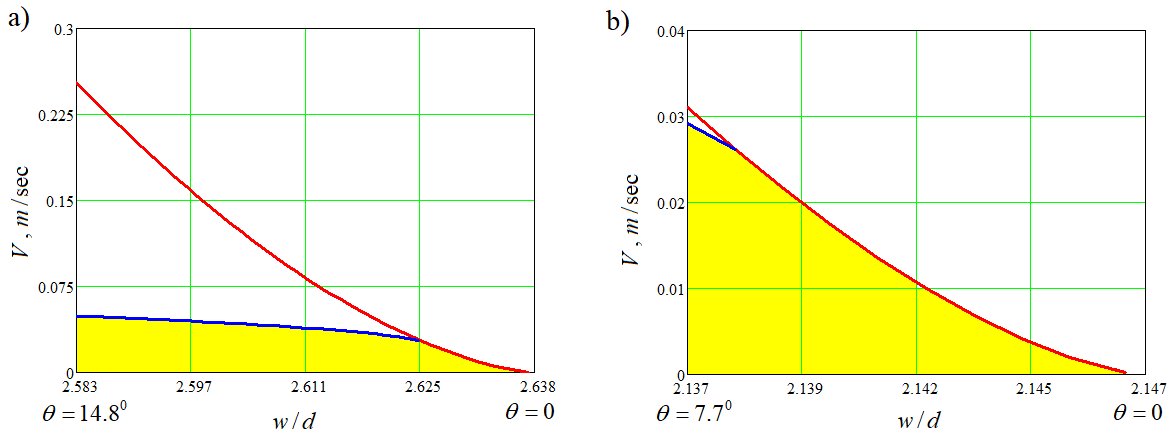}
\caption{Bounds of domains of existence for bridging shown by yellow between the curve of bridge breakdown (blue line) and the curve of instability (red line) for ${\alpha=0.1}$ ({\it a}), ${\alpha=0.2}$ ({\it b}), ${\alpha=0.3}$ ({\it c}) in the variables of velocity $V$ and the fracture width-to-particle size ratio $W/d$. Blue curve is the margin of bridge existence, which depends on the grain strength, and the red line is the bridge existence margin related to the arch instability. Flow deceleration domain filled with yellow is bounded by these curves.}
\label{Fig14}
\end{figure*}

	Note that ceramic proppants, typically used in hydraulic fracturing, are of medium strength. Ceramic proppant is used for the closure (compression) strength up to ${6.9\cdot 10^7}$ Pa. For proppants of the highest strength, the closure stress may reach the values of $10^8$ Pa.
    Consider the grains with the Young's modulus of ${G=2\cdot 10^9}$ Pa and Poisson's ratio ${\nu=0.3}$ and the friction coefficient between the grains in the interval from ${\alpha=0.1}$ to ${\alpha=0.3}$. The wall friction coefficient ${\alpha_w=0.5}$. Grains have the radius $R=5\cdot 10^{-4}$ m. Tensile strength ${\sigma_{fr}=1\cdot 10^7}$ Pa. Fluid viscosity ${\mu=2.2\cdot 10^3}$ Pa${\cdot}$s. The friction coefficient is fixed to be $\alpha=0.2$ and study two configurations of the grains orientation at the angles $\beta=35^{\circ}$ and $\beta=55^{\circ}$. Bounds of the domain of existence for bridging in the plane of coordinates velocity ${V}$ and fracture width-to-grain size ratio are shown in Fig.~\ref{Fig14}. As in the case of loose packing, the bridge disintegration may occur both as a result of the central particle being pushed through the side grains (red curve) and due to grains crushing (blue curve). These curves bound the domains of fluid velocity, where bridging exists (yellow domains in Fig.\ref{Fig14}).  

\subsection{Preparations for numerical implementation}
We will now determine the margins of the interval, where bridging is formed, based on the model proposed above. The bridging "window' is between the values $W_{min}/d$ and $w_{max}/d$. We then determine the level of maximum force that the bridge (an arch) can sustain, or alternatively, above which the central particle will be pushed through. In nondimensional variables, we find, based on Eqs.~(\ref{Eq2}), (\ref{Eq7}), and (\ref{Eq11}):
$$
F_{cr}=\sqrt{\frac{(\cos\theta_{cr}-\cos\theta_0)^3}{\cos^3\theta_{cr}}}(\sin\theta_{cr}+\alpha\cos\theta_{cr}
$$

Here
$$
F_{cr}=P_{cr}\frac{3(1-\nu)}{16GR^2},
$$
where $G$ and $\nu$ are Youngs modulus and Poisson's ratio of the particle material, respectively, and $R$ -- particle radius.

We will now find the minimum fracture width, at which the bridge will be disintegrated under the action of the force exerted from the fluid $P_{fl}$. To simplify the calculations, we introduce the following approximation:
$$
F_{cr}\approx F_{cr}^{int}=\frac{B}{A}\left( \frac{w_{max}}{d}-\frac wd\right)^2,
$$
where $A$ and $B$ are parameters.

\begin{figure}[!htb]
\centering
\includegraphics[width=8cm]{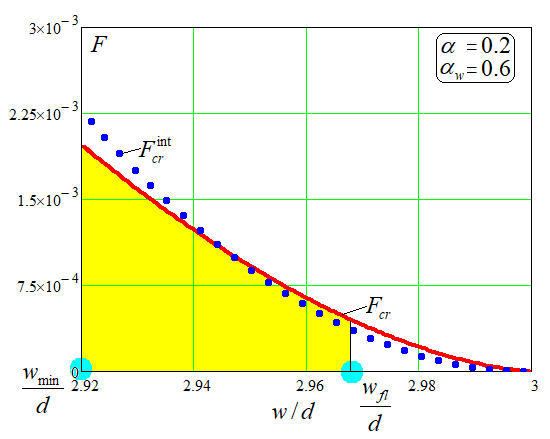}
\caption{Domain of bridging existence (shown as yellow) for $\alpha_w=0.6$ and $\alpha=0.2$. Red line shows the calculated limiting curve, at which the central particle is pushed through. Blue dots are the approximation of the exact solution.}
\label{FigApp1}
\end{figure}

In the case $\alpha=0.2$ and $\alpha_w=0.6$, the parameters are equal to $A=0.076$ and $B=2.5\cdot 10^{-3}$. In Fig.~\ref{FigApp1}, we plot the corresponding curves for $F_{cr}$ and $F_{cr}^{int}$. Then we determine the value of the maximum force $P_{fl}$ based on the given fluid flow velocity $V_{fl}$
$$
P_{fl}=6\pi \mu RV_{fl}
$$

In the nondimensional variables, this expression takes the form:
$$
F_{fl}=P_{fl}\frac{3(1-\nu)}{16GR^2}
$$

Then, the limiting nondimensional width where bridging disintegrated is determined by the formula:
$$
\frac{w_{fl}}{d}=\frac{w_{max}}{d}-\sqrt{F_{fl}\frac AB}
$$

As a result, we obtain the interval $W_{min}/d, w_{fl}/d$m where bridging exists, if $P_{fl}$ does not exceed $P_{cr}=F_{cr}^{int}\frac{16GR^2}{3(1-\nu)}$.

\begin{figure*}[!htb]
\centering
\includegraphics[width=15cm]{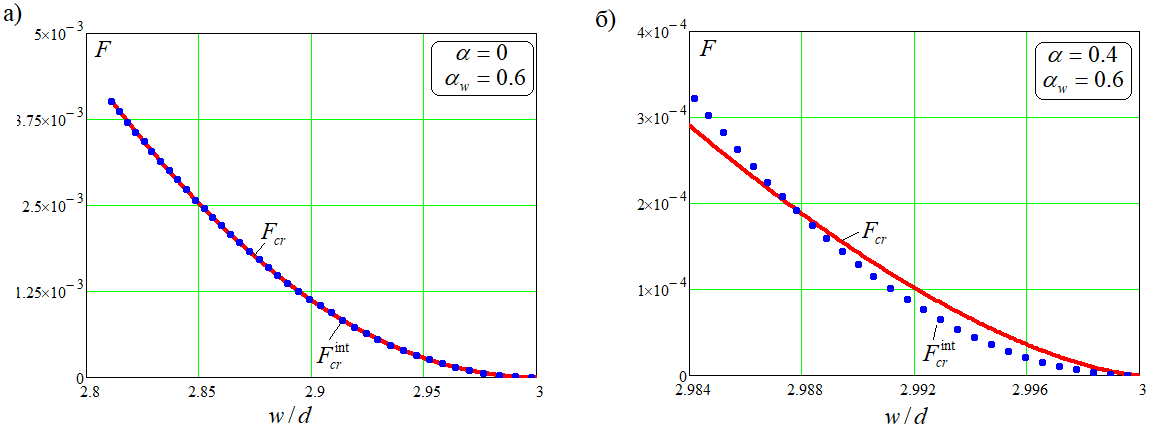}
\caption{Domain of bridging existence (shown as yellow) for $\alpha_w=0.6$ and $\alpha=0.2$. Red line shows the calculated limiting curve, at which the central particle is pushed through. Blue dots are the approximation of the exact solution.}
\label{FigApp2}
\end{figure*}

In Fig.~\ref{FigApp2}, we plot limiting threshold curves (red lines) and appxomating curves (blue dots). Parameters of the approximation are $A=0.189$, $B=4.2\cdot 10^{-3}$ ({\it a}) and $A=0.016$, $B=3.3\cdot 10^{-4}$ ({\it b}).

The results obtained allows us to formulate the bridging criterion in the form, similar to the widely accepted kinematic criterion, but with account for the dynamics effects:
$$
w^* = \mathrm{min}\left[b_{fl},1 +\frac{C_p}{C_{fl}}(b_{fl}-1)\right]d
$$

Here $b_{fl}=\frac{w_{fl}}{d}$ is determined by the formula above, and the corresponding proppant concentration can be obtained from:
$$
C_{fl}=\frac{\pi}{2b_{fl}(a+\sin[\arccos((b_{fl}-1)/2)])}
$$
Consider a numerical example. For $V_{fl}=0.01$ m/s, $G=1\cdot 10^9$ Pa, $\nu=0.25$, $\mu=2.2\cdot 10^3$ Pa s, $R=5\cdot 10^{-4}$ m. Then for $\alpha_w=0.6$ and $\alpha=0.2$, $b_{fl}=2.94$ and $C_{fl}=0.43$.

As a disclaimer, we would like to state that this interval of bridging existence may appear quite model-dependent and should be further validated in proper lab experiments.

%
%
%
%
%
%
%
%
%
%
%
%
%
%
\section{Implementation of the bridging criterion into the 2D width-averaged model of particle transport in a plane channel}
In this section we present the result of numerical simulations of particle transport in a narrow plane channels with at flow conditions similar to hydraulic fracturing jobs. We investigate numerically the effect of dynamic proppant bridging model on the formation of particle accumulation zones.

\subsection{Problem formulation}

In order to give an example on the effect of dynamic bridging on the particle transport, we considered numerical simulations of suspension flow in a narrow vertical channel approximating a hydraulic fracture. The flow is described using the lubrication approximation to Navier-Stokes equations describing the particle-laden flow in a narrow channel approximating the hydraulic fracture. The suspension flow is described using the two-fluid approach with Newtonian incompressible carrier fluid and spherical identical particles described by the volume concentration $C_p$. The system of non-dimensional governing equations describing the particle-laden flow is as follows (see, for example, \cite{Bannikov2018, OsiptsovBoroninDontsov2018, BoroninOsiptsov2010}):

\begin{equation}
\label{PT_1}
\frac{\partial wC_p}{\partial t}+\mathrm{\nabla}\cdot\left(wC_p\mathbf{v}_p\right)=0
\end{equation}
\begin{equation}
\label{PT_2}
\mathrm{\nabla}\cdot\left(\frac{w^3}{12\mu_{m}}\left[\nabla p + \mathrm{Bu}\rho_m\mathbf{e}_y\right]\right)=\frac{\partial w}{\partial t}+(1-C_p)2v_l,
\end{equation}
\begin{equation}
\label{PT_3}
\mathbf{v}_f=-\frac{w^2}{12\mu_{m}}\left(\nabla p + \mathrm{Bu}\rho_m\mathbf{e}_y\right),\,\,\,
\mathbf{v}_p = \mathbf{v}_f + \mathbf{v}_s,\,\,\,\mathbf{v}_s = -v_s\mathbf{e}_2,
\end{equation}
\begin{equation}
\label{PT_4}
v_s=v_{St}f(C_p),\,\,\,f(C_p) = \left( 1 - \frac{C_p}{C_{max}}\right)^5,
\end{equation}
\begin{equation}
\label{PT_5}
\mu_m = \left( 1 - \frac{C_p}{C_{max}}\right)^{-1.89},\,\,\,\rho_m = (1-C_p ) + C_p\zeta_p
\end{equation}
\begin{equation}
\label{PT_6}
\mathrm{Bu}=\frac{\rho gw_0^2}{\mu U},\,\,\,\zeta_p=\frac{\rho_p}{\rho},\,\,\,v_{St}=\frac{2a^2\rho g (\zeta_p - 1)}{9\mu U}.
\end{equation}

Here, $w(x,y)$ is the channel width (a prescribed function of coordinates); $v_l$ fluid leak-off velocity through the channel walls; $\rho_m$ and $\mu_m$ are the density and viscosity of the suspension, respectively. 

The system (\ref{PT_1}), (\ref{PT_2}) is formulated in a Cartesian coordinate system $\mathrm{O}xy$ with the horizontal $x$-axis and vertical $y$-axis, the origin $O$ is located at the fracture middle plane. The following scales are used: fluid velocity at the fracture inlet $U$, fracture length $L$ and width $w_0$ (so that $d/L \ll 1$). Functions $f(C_p)$ (\ref{PT_4}) and $\mu_m(C_p)$ (\ref{PT_5}) determine the closure relations for suspension viscosity and particle settling velocity, respectively. The equations contain the following non-dimensional parameter determining the flow: Buoyancy number $\mathrm{Bu}$, particle-to-fluid substance density ratio $\zeta_p$ and dimensionless particle settling velocity $v_{St}$.

Boundary conditions include zero flux through the horizontal boundaries of the computational domain ($y = 0$ and $H/L$, where $H$ is the channel height), a prescribed velocity at inlet ($x = 0$) and hydrostatic pressure at outlet ($x = 1$) describing the smooth boundary conditions:
\begin{equation}
\label{13}
p(1,y)=-\mathrm{Bu}\int\limits_0^y\rho_m(1,s)\mathrm{d}s
\end{equation}

\subsection{Input parameters and channel configuration}
In the numerical simulations of particle transport in a channel, we used the following input parameters typical of real hydraulic fracturing jobs: channel length $L$ and height $H$ are $100\,m$ and $20\,m$, respectively; channel width scale $w_0$ is $5\cdot10^{-3}\,m$; velocity scale $U$ is set to $0.025$; inlet volumetric flux is $2\cdot10^{-3}\,m^3/s$; fluid viscosity $\mu$ is $3\cdot10^{-3}\,Pa\cdot$s and density $\rho$ is $10^3\,kg/m^3$; particle radius $a$ is $5\cdot10^{-4}\,m$ and substance density $\rho_p$ is $10^3$ (neutrally-buoyant) and  $2.6\cdot10^3\,kg/m^3$ (sand); particle volume concentration at the inlet $C_{in}$ is $0.05$. Fluid leakoff is set to zero (impermeable walls). Though leak-off through fracture walls is very important for the global coupled model~\cite{Mikhailov}, in this example we switch it off to be able to separate the effects of pure bridging and suspension packing due to dehydration. The value of inlet volumetric flux is chosen to match velocities inside the channel with that in a typical hydraulic fracture. The channel is filled initially with the carrier fluid and particle-laden suspension is injected during a certain time period with the particle volume concentration $C_p = C_{in}$.

The following channels with static walls are considered:
\begin{enumerate}
\item Linear channel \begin{equation}\label{LinearChannel} w(x) = (w_1 - 1)x + 1\end{equation}
\item Elliptic channel \begin{equation}\label{EllipticChannel}w(x,y) = ([w_1 - 1]x + 1)\left(w_2 + \sqrt{1 - \left[\frac{2y-h}{h}\right]^2}\right)\end{equation}
\end{enumerate}
Here, $w_1=0.3$, $w_2 = 0.65$ and $h=H/L$. The aperture of both channels decreases linearly with an increase in the horizontal coordinate $x$. The first channel has a square vertical cross-section, while the second channel has the cross-section with the width varying with vertical coordinate $y$ (close to ellipse).

In order to illustrate the effect of fluid velocity on proppant bridging, we considered an artificial bridging criterion in the following form:
\begin{eqnarray}
\label{Bridg_Sim}
B = 1&\textrm{if}& w^* > 2.5\,d \\ \nonumber
&\textrm{or}&2.5 < w^* < 3\,\, \textrm{and}\,\, \mathbf{v}^*_f>v^*_{crit} \\ \nonumber
B = 0& & \textrm{otherwise}
\end{eqnarray}
Here, $B$ is the bridging function ($\mathbf{v}_p = B\mathbf{v}_f$), so that $B = 0$ corresponds to bridged particles, while $B = 1$ corresponds to moving particles; $\mathbf{v}^*_f$ is dimensional fluid velocity; $b = 2.5$; $d=2a$ is the particle diameter; $w^*=w_0\,w$ is the dimensional channel aperture; $v^*_{crit}$ is dimensional critical velocity determining instability of a bridged proppant determined as follows: 
\begin{equation}
\label{V_crit}
v^*_{crit} = v^*_{crit,0}\left(1-2\left[\frac{w^*}{d} - b\right]\right)
\end{equation}

In numerical simulations we considered $v^*_{crit, 0}=0.05\,m/s$, which is of the same order of magnitude as the maximum velocities of the proppant pack stability obtained above (see Figs.~(\ref{Fig7}), (\ref{Fig14})).  The dynamic bridging condition (\ref{Bridg_Sim}) is compared against the standard condition mentioned in the introduction:
\begin{eqnarray}
\label{Bridg_Std}
B = 1&\textrm{if}& w^* > 2.5\,d \\ \nonumber
B = 0& & \textrm{otherwise}
\end{eqnarray}

While the simulations are aimed at analyzing the proppant bridging, we neglected the effect of particle settling and set the particle settling velocity to zero ($v_s = 0$ in (\ref{PT_4})). Therefore, particles only modify the suspension properties (viscosity $\mu_m$ and density $\rho_m$ according to (\ref{PT_5})) and provoke suspension slumping due to density contrast between the suspension and the particle-free fluid. Note that since the particle velocity slip is neglected and the carrier fluid is incompressible, particle concentration redistribution occurs only in the vicinity of zone occupied by bridged proppant. There are two opposing particle redistribution mechanisms, namely, i) formation of bridged zones (when bridging criterion is met) leading to local particle accumulation, and ii) mobilization of bridged proppant due to changes in local fluid velocity. The latter mechanism is expected to occur regardless the proppant bridging model (\ref{Bridg_Sim}) or (\ref{Bridg_Std}), in particular, when the flow changes direction in the vicinity of bridged proppant and particles move towards the direction, where the bridging criterion is not met.

\subsection{Numerical implementation}
Numerical solution to Eqs.~(\ref{PT_1}), (\ref{PT_2}) is carried out using the finite-difference method and staggered uniform rectangular mesh. IMPES strategy for splitting the governing equations is considered. At each time step, the linear elliptic pressure equation (\ref{PT_2}), in which the coefficients are approximated at the previous time step, is solved (using a preconditioned BiCGStab solver). Then the velocity field $\mathbf{v}_f$ is updated, and the particle volume concentration $C_p$ is advected according to the numerical solution of Eq.~(\ref{PT_1}) carried out using second-order explicit TVD scheme. 

In the simulations we fixed the mesh resolution to $301\times 101$ nodes in horizontal and vertical directions, respectively. In all test cases considered we tracked the particle mass balance to preserve the particle mass conservation, the relative proppant mass balance error was below $10^{-3}$. The numerical algorithm is implemented in C++ programming language by Skoltech team. 

\subsection{Results of numerical simulations}
\begin{figure*}[!htb]
\centering
\includegraphics[width=9cm]{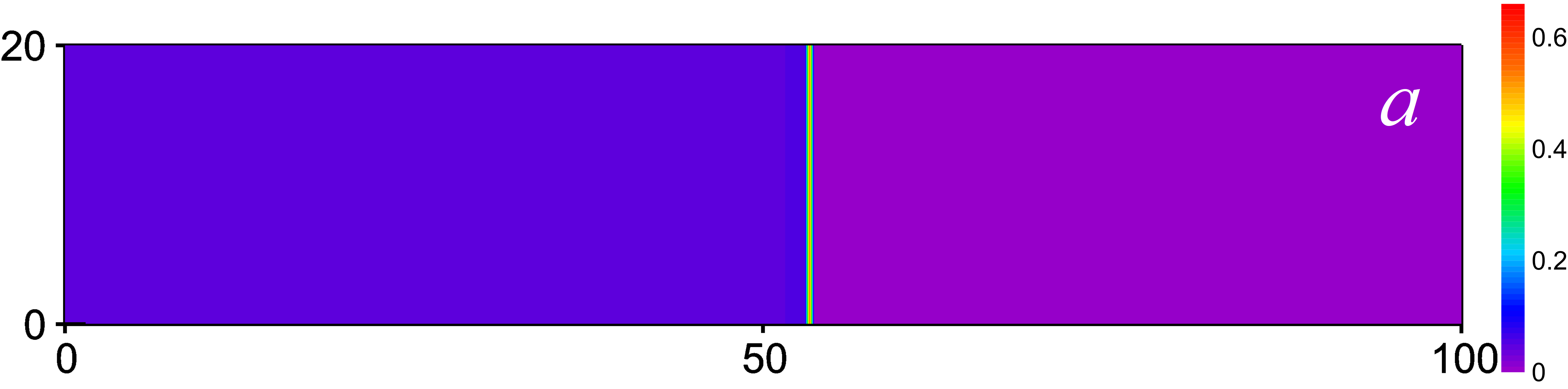}\hspace{0.25cm}
\includegraphics[width=9cm]{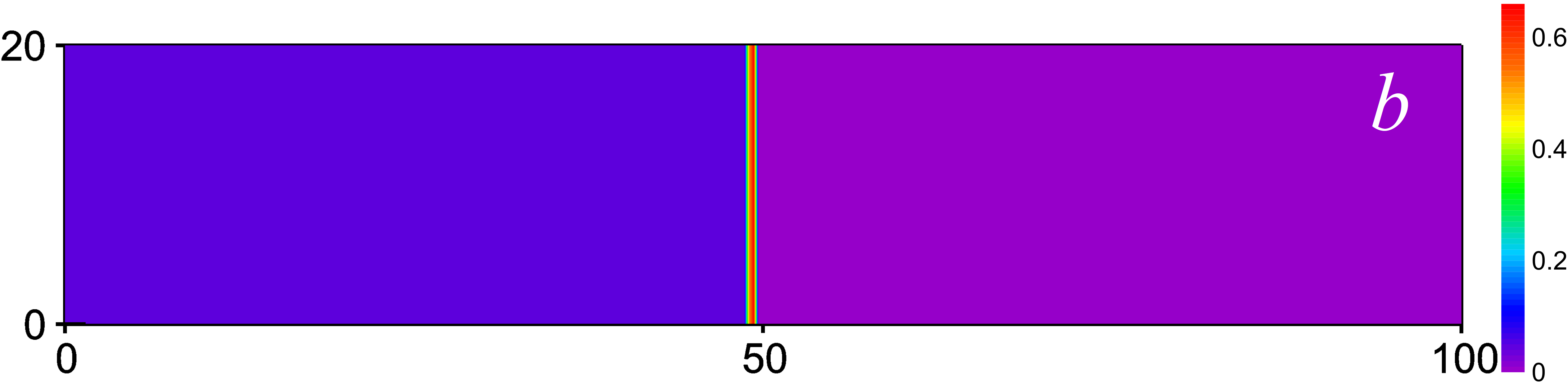}

\includegraphics[width=9cm]{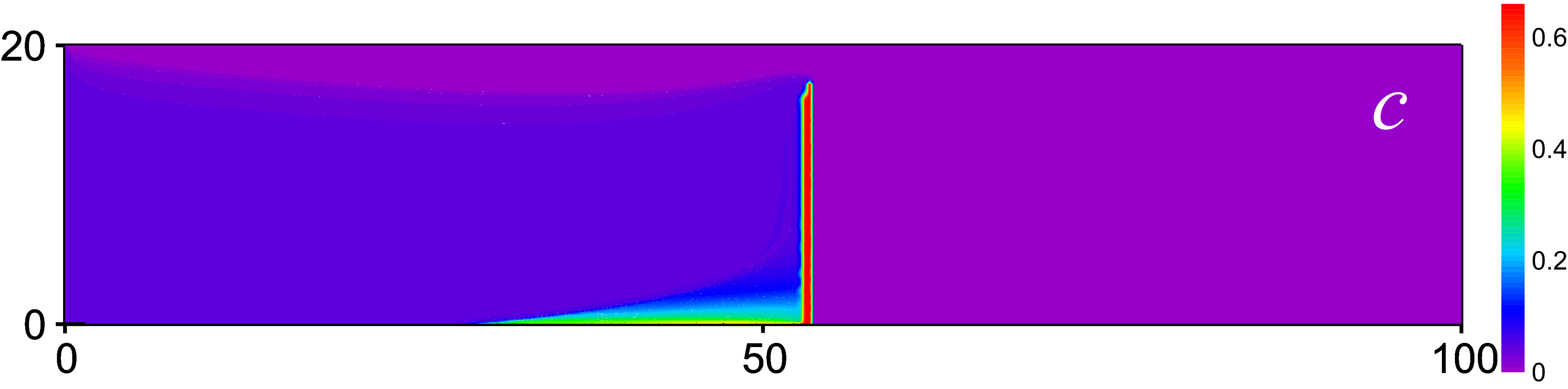}\hspace{0.25cm}
\includegraphics[width=9cm]{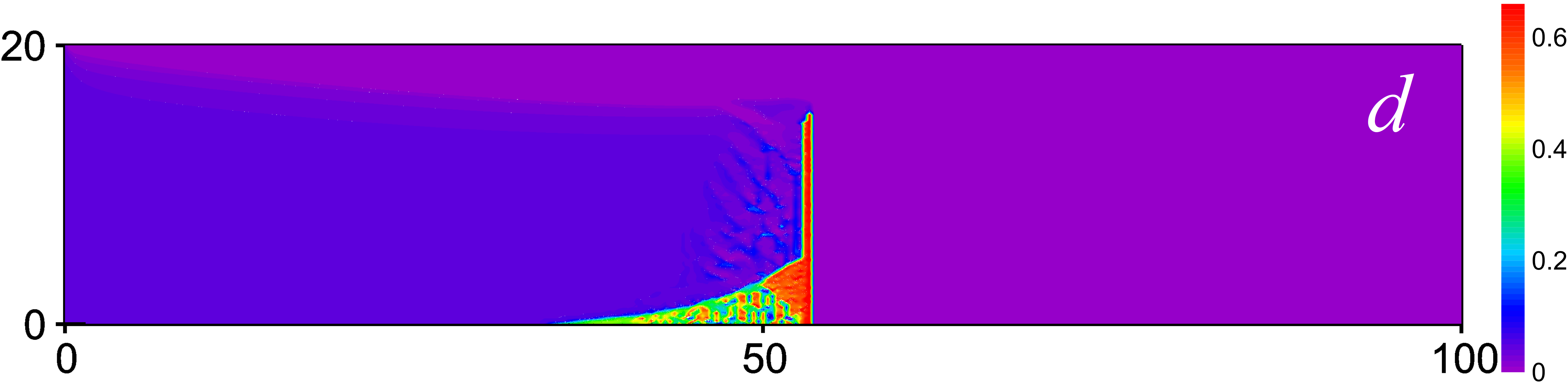}
\caption{Distribution of the particle volume concentration in the channel with a square cross-section (\ref{LinearChannel}) at the end of injection (see Table~\ref{Tab1} for injection times). Proppant bridging is determined by the standard proppant bridging model (\ref{Bridg_Std}) (\textit{a}, \textit{c}) and improved dynamic bridging model (\ref{Bridg_Sim}) (\textit{b}, \textit{d}). Plots (\textit{a}, \textit{b}) show results for neutrally-buoyant particles, while (\textit{c}, \textit{d}) correspond to sand particles.}
\label{Fig15}
\end{figure*}

\begin{figure*}[!htb]
\centering
\includegraphics[width=9cm]{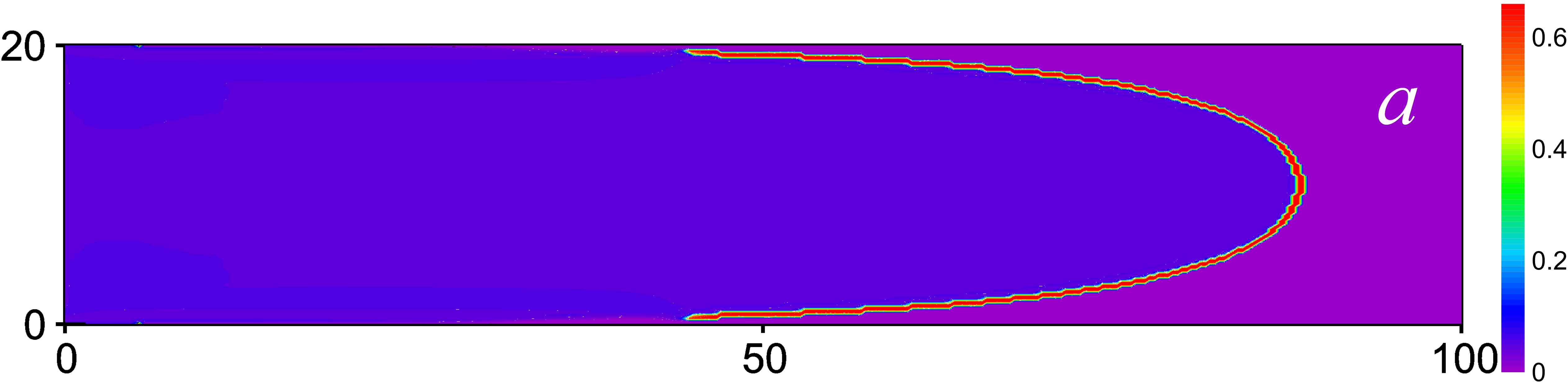}\hspace{0.25cm}
\includegraphics[width=9cm]{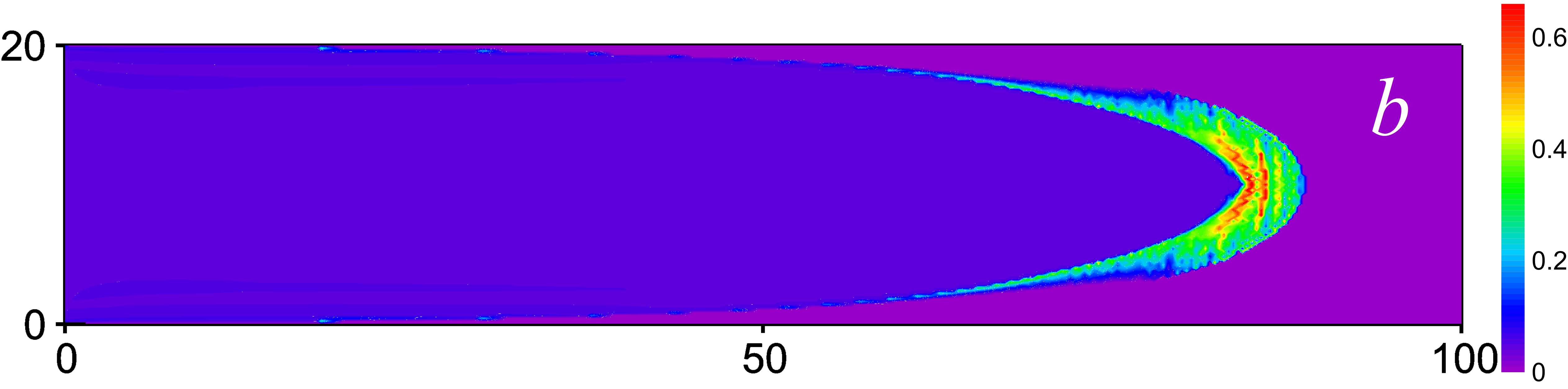}

\includegraphics[width=9cm]{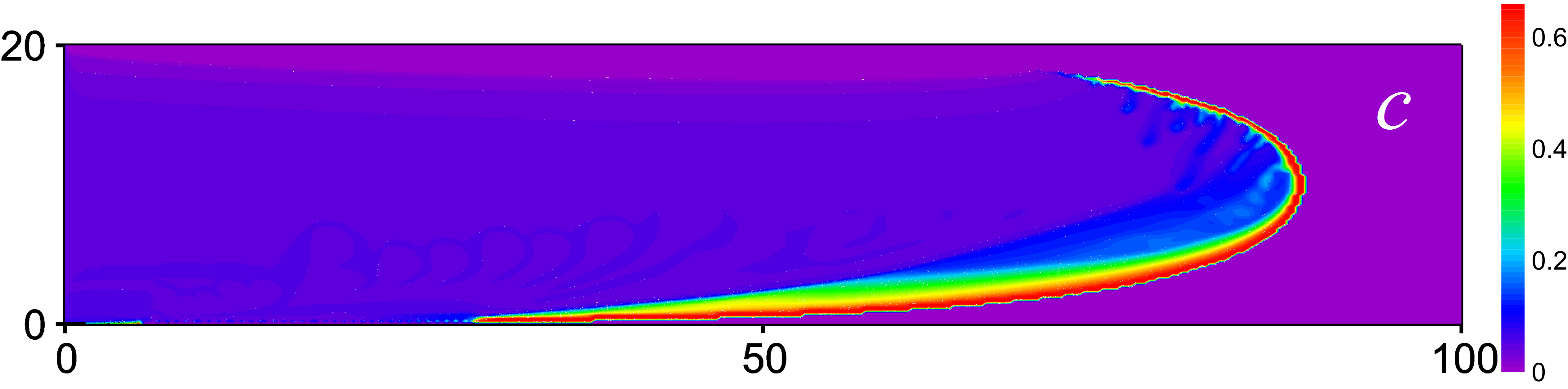}\hspace{0.25cm}
\includegraphics[width=9cm]{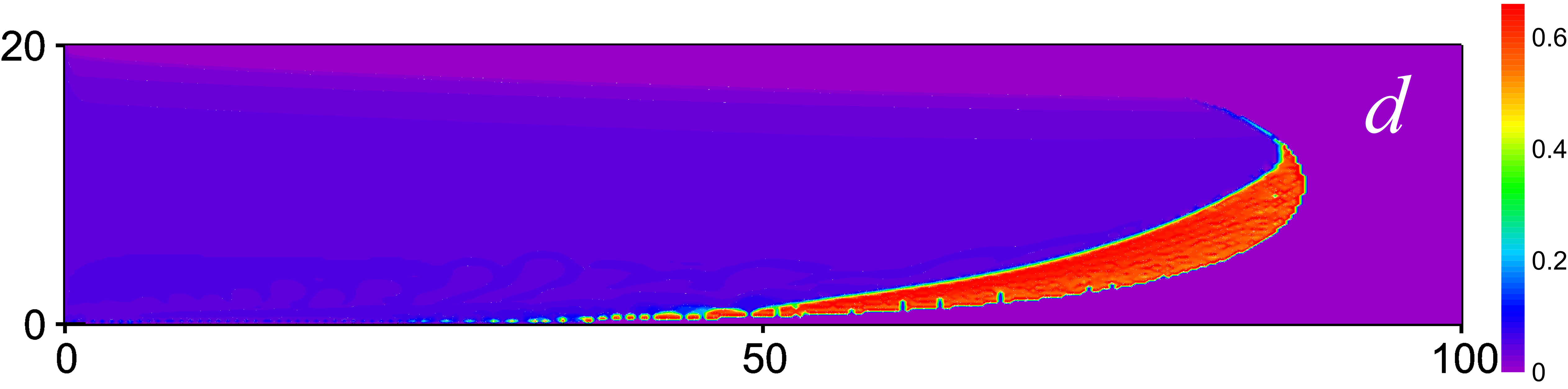}
\caption{Distribution of the particle volume concentration in the channel with ``elliptic'' cross-section (\ref{EllipticChannel}) at the end of injection (see Table~\ref{Tab1} for injection times). Proppant bridging is determined by the standard proppant bridging model (\ref{Bridg_Std}) (\textit{a}, \textit{c}) and improved dynamic bridging model (\ref{Bridg_Sim}) (\textit{b}, \textit{d}). Plots (\textit{a}, \textit{b}) show results for neutrally-buoyant particles, while (\textit{c}, \textit{d}) correspond to sand particles.}
\label{Fig16}
\end{figure*}

We obtained that the particle concentration distribution in both channel types considered is affected significantly by the proppant bridging model (Figs.~\ref{Fig15}, \ref{Fig16}). The problem of hydraulic fracture propagation involves not only particle transport, but also geomechanics of rock fracturing, in which the fluid pressure inside a hydraulic fracture plays a key role. Therefore for all test cases considered, we provided the dimensionless pressure drops along the central line of the slots $y=h/2$ (see Table~\ref{Tab1}). The pressure distribution along the channel allows us to evaluate qualitatively the effect of proppant bridging criterion on the fracture propagation. Quantitative effect of dynamic bridging on hydraulic fracture propagation and the final fracture shape could be studied only in the framework of coupled fluid mechanics and geomechanics models (e.g., Pseudo 3D or Planar 3D).

In the case of the square channel (\ref{LinearChannel}) and neutrally-buoyant particles (Fig.~\ref{Fig15}(\textit{a}, \textit{b})) dynamic bridging criterion provokes earlier particle bridging as compared to the standard one. This is justified by the location of a layer of packed proppant and the pressure drop along the channel, which is larger in case of dynamic bridging. The presence of slumping (sand particles) provides vise-a-versa result: the height of dynamically bridged proppant is less than that of a proppant bridged due to static criterion (Fig.~\ref{Fig15}(\textit{c}, \textit{d})). This results in about 20\% pressure drop difference between these two cases (see Table~\ref{Tab1}).

Dynamic proppant bridging results in much larger pressure drop along the channel with ``elliptic'' cross-section (\ref{EllipticChannel}) for both particle types (see Fig.~\ref{Fig16}). Note that the channel is almost completely bridged in case of neutrally-buoyant particles at $t = 65\,min$, while dynamic bridging provides significantly different shape of the zone of immobilized proppant corresponding to much smaller pressure drop along the channel.

By analyzing the distributions of sand particles in channels shown in Figs.~\ref{Fig15}(\textit{c}, \textit{d}), \ref{Fig16}(\textit{c}, \textit{d}), one can see that there are zones with significant area where the particle concentration is between the inlet concentration $C_{in} = 0.05$ and the maximum packed concentration $C_{max} = 0.65$. We claim that that this is a result of the second particle concentration redistribution mechanism discussed above at the end of section 3.2 - proppant is mobilized due to local change is flow direction. This is confirmed by numerical simulations carried out at finer mesh: the shape of ``diffusion'' areas do not change significantly and the local velocities indeed change sign in these zones.

\begin{table*}
\begin{center}
\begin{tabular}{|l|l|l|l|l|}
\hline
Channel cross-section & Particle type & Injection time (min) & 
$\Delta p$, standard bridging (\ref{Bridg_Std}) & $\Delta p$, dynamic bridging (\ref{Bridg_Sim}) \\
\hline
Square & \begin{tabular}{l} Neutrally-buoyant \\ Sand \end{tabular} &  \begin{tabular}{l} 30.5 \\ 30 \end{tabular} & \begin{tabular}{l} 89.15 \\ 120.9 \end{tabular} & \begin{tabular}{l} 93.53 \\ 103.9 \end{tabular} \\ \hline
Elliptic & \begin{tabular}{l} Neutrally-buoyant \\ Sand \end{tabular} & \begin{tabular}{l} 65 \\ 74 \end{tabular}  & \begin{tabular}{l} 327.96 \\ 65.9 \end{tabular} & \begin{tabular}{l} 41.25 \\ 40.72 \end{tabular} \\ \hline
\end{tabular}
\label{Tab1}
\caption{Calculated dimensionless pressure drop along the center-line of the channel $\Delta p$ (at $y = h/2$) for the two channel types and the two particle types considered. The pressure scale is $\mu U L / w_0^2$.}
\end{center}
\end{table*}

\section{Discussion}

In this work, we made an effort to advance the theory of bridging of proppant grains in a hydraulic fracture from  semi-empirical kinematics ($w=2.5d$) to a dynamic criterion based on the consideration of forces exerted on a single particle. Existing models which are implemented into commercial simulators use the simplified geometric condition for bridging, which dates back to the work~\cite{GruesbeckCollins} on particle bridging on circular perforations. Generalizations of that criterion included the effect of particle volume concentration, but did not re-visit the basics, namely, the geometry (which planar channel not a circular tube) and the dynamics of the process.

Now, at the same time there is a variety of ongoing studies on bridging that are based on lab experiments, in particular,~\cite{BarreeConway2001}, and a more recent work~\cite{Ray}. These studies report particular investigations on proppant bridging, packing and holdup in particular geometries. The studies discuss the validity of the geometric criterion and suggest that for some proppant bridging occurs only when the slot spacing is comparable with the grain diameter. There is also a chain of papers that employ a coupled CFD-DEM/FEM approach to direct numerical simulation of particle bridging at constrictions (e.g.,~\cite{Sharma2016}). 

We suggest that the theoretical approach proposed in the present paper could be now calibrated and validated in properly designed lab experiments in order to close the research loop and conclude in a lumped dynamic criterion, which could then be implemented into commercial simulators of hydraulic fracturing based on next generation coupled models, such as, for example,~\cite{Dontsov2014, Dontsov2015, Bannikov2018, OsiptsovBoroninDontsov2018, Erofeev2018, Golovin2018}.

\section{Conclusions}

We studied in detail the phenomenon of dynamic bridging of proppant particles during suspension flow in a hydraulic fracture. In contrast to existing kinematic criteria for bridging formulated empirically ($W/d=2.5$), the effort is made to come up with the criterion of bridging formation and disintegration based on detailed consideration of the dynamics of the process by means of direct numerical simulation and solid mechanics considerations. We considered both loose packing and close packing configurations around 3 particles across the fracture.

We have formulated the dynamic bridging criterion in the plane of the two key variables: the fracture width-to-particle diameter ratio $W/d$ and fluid velocity $V$, depending on the parameters characterizing frictional and elastic interaction in between the particles and between the grains and the fracture walls (inter-particle friction coefficient $\alpha$, particle-wall friction coefficient $\alpha_w$, and elastic moduli). 

It is demonstrated that there is an interval in terms of the fluid flow velocity which favors bridging formation $V_1\le V \le V_2$. For $V<V_1$, the bridge forms but slips on the walls. Within the interval $V_1\le V \le V_2$, the particles in the bridge are all in contact, and the hydrodynamics force exerted on the central sphere translates to the wall friction, which is sufficient to keep the bridge immovable across the fracture. When $V \ge V_2$, the load on the central particle is overcritical for the central particle to move through, between the two side grains. 

Numerical simulations of the particle transport in a plane channel carried out in the framework of 2D particle transport model showed that the dynamic proppant bridging criterion provide a significant effect on the shape and total area of bridged zones. Also dynamic bridging in most of the test cases considered provides much smaller pressure drops along the channel, which could potentially provide a significant impact on the final shape of a hydraulic fracture. Accurate theoretical study of the effect of dynamic proppant bridging on the hydraulic fracturing process requires simulations run in the framework of coupled geomechanics and fluid mechanics models and is a subject of our future work.

\section*{Acknowledgements}
{This work received financial support from the Ministry of Education and Science of the Russian Federation (project №14.581.21.0027, unique identifier RFMEFI58117X0027).

The authors are grateful to the management of LLC "Gazpromneft-STC'' and LLC "MIPT Center for Engineering and Technology'' for financial support of this work.

Startup funds of Skolkovo Institute of Science and Technology are gratefully acknowledged by A.A. Osiptsov.}

\bibliographystyle{elsarticle-num}
\bibliography{Drill-bit-cutters.bbl}
\end{document}